\documentclass[epj]{webofc}
\usepackage[varg]{txfonts}   
\usepackage{lineno}

\begin{document}

\title{Multimessenger Probes of High-energy Sources}
%
%

\author{Dafne Guetta\inst{1}\fnsep\thanks{\email{dafneguetta@braude.ac.il}}}

\institute{ORT Braude College, Karmiel, Israel}

\abstract{Multimessenger observations may hold the key to learn about the most energetic sources in the universe. The recent construction of large scale observatories opened new possibilities in testing non thermal cosmic processes with alternative probes, such as high energy neutrinos and gravitational waves.
We propose to combine information from gravitational wave detections, neutrino observations and electromagnetic signals to obtain a comprehensive picture of some of the most extreme cosmic processes. Gravitational waves are indicative of source dynamics, such as the formation, evolution and interaction of compact objects. These compact objects can play an important role in astrophysical particle acceleration, and are interesting candidates for neutrino and in general high-energy astroparticle studies.
In particular we will concentrate on the most promising gravitational wave emitter sources: compact stellar remnants. The merger of binary black holes, binary neutron stars or black hole-neutron star binaries are abundant gravitational wave sources and will likely make up the majority of detections. However, stellar core collapse with rapidly rotating core may also be significant gravitational wave emitter, while slower rotating cores may be detectable only at closer distances.
The joint detection of gravitational waves and neutrinos from these sources will probe the physics of the sources and will be a smoking gun of the presence of hadrons in these objects which is still an open question.
Conversely, the non-detection of  neutrinos or gravitational waves from these sources will be fundamental to constrain the hadronic content.}
\maketitle
\section{Introduction}

With the contemporary operation of the IceCube and Antares, Advanced LIGO and Advanced Virgo facilities, we are now able to observe the universe using two new, distinct astrophysical messengers. In addition to photons from radio waves to gamma-rays, we can now, for the first time, simultaneously observe the sky with neutrinos and gravitational waves (GWs). 

This new era of multi-messenger astrophysics offers a unique view of the universe and provide powerful insights into the workings of some of the most energetic and enigmatic objects in the cosmos.

High-energy cosmic neutrinos observed by the IceCube experiment reveal a deep view of the universe at energies where the sky is opaque to photons. With more than one thousand times the energy of the most energetic neutrinos produced with earthbound accelerators, cosmic neutrinos collected by IceCube also also exceed by a factor of one billion the energy of the neutrinos detected from the supernova explosion in the Large Magellanic Cloud detected in February 1987, the only astrophysical neutrinos observed from outside the solar system prior to IceCube’s breakthrough.
An immediate inference made about the large neutrino flux observed by IceCube, which is predominantly of extragalactic origin, is that the total energy density of high-energy neutrinos in the universe is similar to that of gamma-rays. This is worthy of a closer look. Astrophysical neutrinos are mostly the decay products of pions. Protons accelerated in regions of high magnetic fields near neutron stars or supermassimve black holes (like in jet og Active Galactic Nuclei) may interact with the radiation or dust surrounding them to produce these pions and kaons that decay into neutrinos. 

Together with charged pions generating neutrinos, neutral pions are produce, which promptly decay into two gamma-rays. These gamma-rays in the matching 10 to 10,000\,TeV energy range of IceCube neutrinos can not reach Earth, as they interact with extragalactic (mostly microwave) photons. This interaction triggers a cascade process that leads to photons in the GeV-TeV energy range, which are detected.


The matching energy densities of the extragalactic gamma-ray flux detected by Fermi and the high-energy neutrino flux measured by IceCube would be a natural consequence of having been originated from  common sources. However, the majority of the extragalactic, unresolved gamma-ray flux is thought to originate from blazars, which alone cannot explain the observed cosmic neutrinos\cite{2015arXiv151100688B}. 
A possible explanation that does not violate the gamma-ray flux limit nor the source emission constraints is that neutrino sources or their environment absorb gamma-rays; i.e. they are electromagnetically hidden \cite{2016PhRvL.116g1101M}. 


Neutrino astronomy represents a unique tool within multi-messenger astrophysics to probe the most extreme cosmic processes. IceCube plans to extend our reach for neutrino sources by instrumenting 10 km$^3$ of glacial ice at the South Pole, improving IceCube’s sensitive volume by an order of magnitude \cite{2014arXiv1412.5106I}. The new facility will increase the event rates for the highest energy neutrinos  from thens to hundreds per year over several years. 

On the other side of the world from IceCube is the ANTARES telescope, in the Mediterranean sea. ANTARES is currently the only deep sea high energy neutrino telescope that is operating in the Northern hemisphere. 
The telescope covers an area of about 0.1 km$^2$ on the sea bed, at a depth of 2475\,m, 40\,km off the coast of Toulon, France. ANTARES is planned to be followed by a multi-kilometer detector in the Mediterranean sea called KM3NeT in the next few years. KM3NeT is the future generation of under water neutrino telescopes. The infrastructure will consist of three so-called building blocks, each made of 115 strings of 18 optical modules, that have 31 photo-multiplier tubes each. KM3Net is made of KM3NeT/ARCA (Toulon, France) and KM3NeT/ORCA (Capo Passero, Sicily)\cite{2006NIMPA.567..457K}.   
The realization of next generation high energy detectors like CTA for TeV photons, KM3Net and IceCube-Gen2 for higher energy neutrinos and the improving sensitivity of GW detectors will open a new era in multi-messenger astrophysics that we propose to exploit. 

Beyond neutrinos, another major pillar of multi-messenger probes are GWs. Advanced LIGO \cite{2015CQGra..32g4001T} recently made the first direct detection of GWs \cite{2016PhRvL.116f1102A}. On September 14$^{th}$, 2015, the LIGO detectors recorded a signal that was soon reconstructed to have come from the merger of two black holes, each with $\sim 30$\,M$_{\odot}$, about 1.3 billion light years away. This accomplishment was not only the beginning of GW observations that can probe binary black holes system and their gravitational interaction in a unprecendented ways; it also started a new chapter for multi-messenger astronomy.

Combining information from GW detections with electromagnetic and neutrino observations allows us to gain a better understanding of some of the most extreme cosmic processes \cite{2013RvMP...85.1401A}. GWs are indicative of source dynamics, such as the formation, evolution and interaction of compact objects. These compact objects are anticipated to play an important role in astrophysical particle acceleration and high-energy emission, making a direct link between the GW observation and high-energy neutrino emissionand in general high-energy astroparticle studies.

\begin{figure}
\begin{center}
\resizebox{0.95\textwidth}{!}{\includegraphics{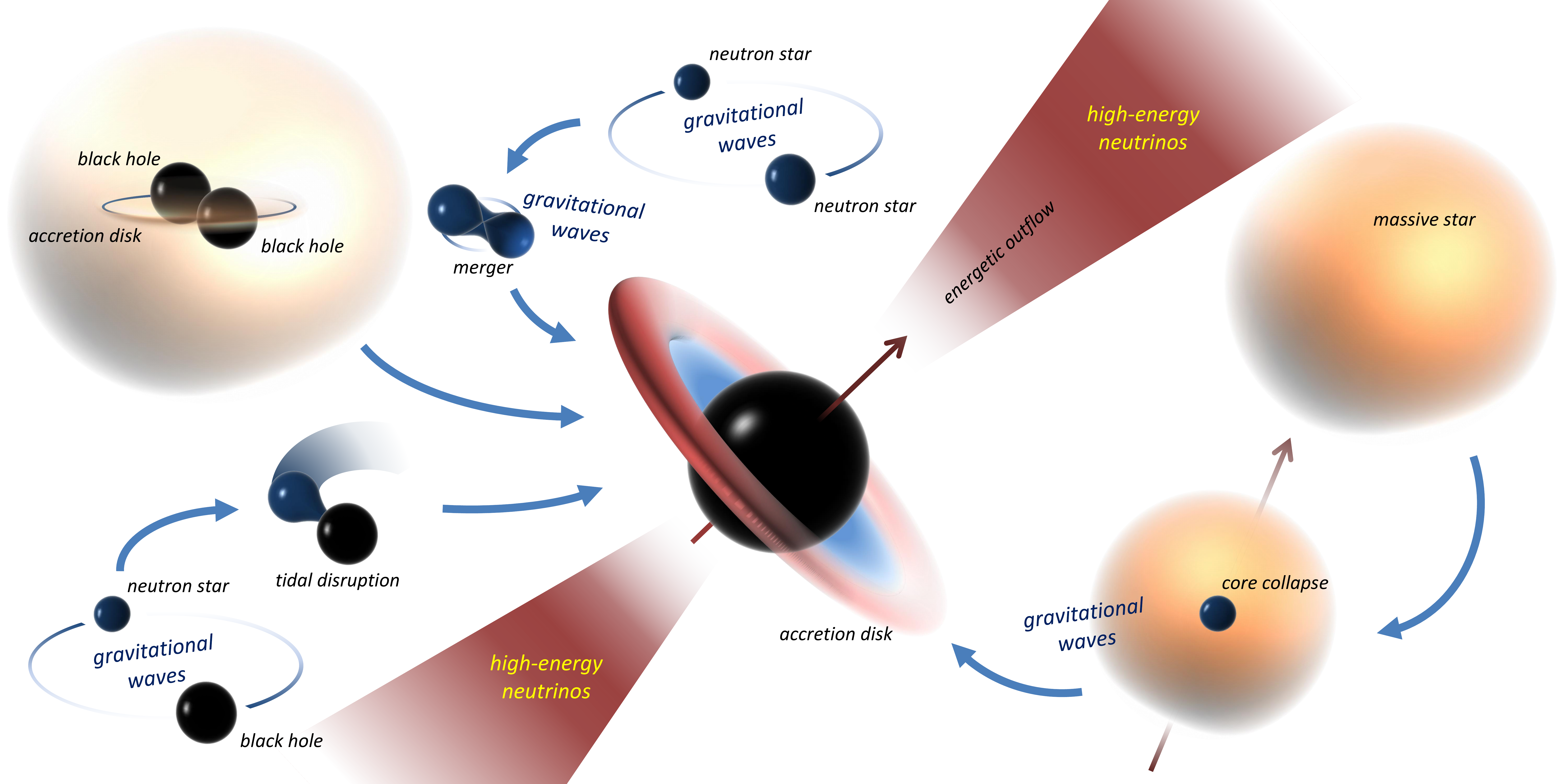}}
\end{center}
\caption{Illustration of multi-messenger sources of GWs and high-energy neutrinos.}
\label{fig:illustration}
\end{figure}

The most promising GW emission is related to compact stellar remnants. The merger of binaries of black holes and neutron stars are abundant GW sources and will likely make up the majority of detections \cite{2016LRR....19....1A}. 
Stellar core collapse with a rapidly rotating core may also be a significant GW emitter \cite{2013CQGra..30l3001B,2016ApJ...818...94K,2015ApJ...805...82M}, while slower rotating cores may be detectable at closer distances \cite{2009CQGra..26f3001O}. If the core collapse results in a protoneutron star, fallback accretion can significantly increase the angular momentum thus resulting in increased GW production (along with electromagnetic emission), improving detection prospects \cite{2012ApJ...761...63P}. 
Other GW sources include rotating neutron stars \cite{1998ApJ...501L..89B}, and plausibly magnetar flares \cite{2008PhRvL.101u1102A,2011PhRvD..83j4014C,2013PhRvD..87j3008M}. See Fig. \ref{fig:illustration} for an illustration of some of the main GW sources that may also produce high-energy neutrino emission.

With the onset of GW observations, there has been a significant effort to search for electromagnetic and neutrino emission from GW sources \cite{2013APh....45...56S}. Started during the operation of initial LIGO and Virgo \cite{2007AAS...211.9903P,2008CQGra..25r4034K}, electromagnetic follow-up efforts now include a large number of partner observatories from radio to gamma-rays \cite{2016ApJ...826L..13A}. A significant number of these telescopes searched for counterparts of the first observed GW, GW150914, but no obvious one was found \cite{2016ApJ...826L..13A}.

Joint GW+neutrino searches started with initial GW detector operations as well. These searches adopted a $\pm500$\,s time window around the gravitational wave event \cite{2011APh....35....1B}, and most of them were based on the same baseline GW+neutrino search technique \cite{2012PhRvD..85j3004B}. The first observational constraints for joint sources were derived using initial LIGO-Virgo and the IceCube detector \cite{2011PhRvL.107y1101B}, which was soon followed by searches with ANTARES \cite{2013JCAP...06..008A} and IceCube \cite{2014PhRvD..90j2002A}. Most recently, high-energy neutrino searches were carried out for the first GW detection, GW150914, with the ANTARES and IceCube detectors \cite{2016PhRvD..93l2010A} and the Pierre Auger Observatory \cite{2016arXiv160807378T}, while the KamLAND detector \cite{2016arXiv160607155K} and IceCube \cite{2016PhRvD..93l2010A} searched for MeV neutrino counterpart. No significant temporally and directionally coincident neutrinos were found by these searches.

\section{Multi-messenger Detection of IceCube Alert IceCube-170922A}

IceCube detects muon neutrinos above a threshold of approximately 10\,GeV, resulting in one well-reconstructed upgoing muon track event every few minutes. These events are dominated by low-energy atmospheric neutrinos. In 2016, IceCube installed an online filter that selects from this sample, in real time, very high energy events that are potentially cosmic in origin, typically having a probability greater than 50\% of coming from outside the earth's atmosphere. These are reconstructed within less than one minute, and the energy and arrival direction of the neutrino are communicated to the Gamma-ray Coordinate Network (GCN) for follow-up by astronomical telescopes. The tenth such alert \cite{ic170922GCB}, IceCube-170922A, on September 22, 2017, reported a well-reconstructed muon neutrino with an energy of 290\,TeV and, therefore, with a high probability of originating in an astronomical source.

What makes this alert special is that, for the first time, telescopes detected enhanced gamma-ray activity from a flaring blazar aligned with the cosmic neutrino to within less than 0.06 degree. The source is a known blazar, TXS0506+056, and its redshift has been subsequently measured to be $z\sim0.34$ \cite{2018ApJ...854L..32P}. Originally detected by NASA's Fermi \cite{fermi-ic170922GCN} and Swift \cite{swiftXRT-ic170922GCN} satellite telescopes, the alert was followed up by the MAGIC air Cherenkov telescope \cite{MAGIC-ic170922GCN}. MAGIC detected the emission of gamma rays with energies exceeding 400\,GeV. With a redshift of 0.34, we can conclude that the source is a TeV blazar. Several other telescopes subsequently observed the flaring blazar.

It is important to realize that nearby blazars like the Markarian sources are at a redshift that is ten times smaller, and therefore TXS0506+056, with a similar flux despite the greater distance,is between the 3
 that multiple attempts have not found a correlation between the arrival directions of cosmic neutrinos previously observed by IceCube and the various Fermi blazar catalogues that are dominated by ``vanilla" nearby sources; we can infer to this flare peculiar luminosy and duration as well as a particular hadron activity.

Given where to look, IceCube searched its archival neutrino data up to and including October 2017, for evidence of neutrino emission at the location of TXS0506+056. When searching the sky for point sources of neutrinos, two analyses have been routinely performed: one looking for a steady emission and one that searches for flares over a variety of timescales. Evidence was found for a spectacular burst of 14 high-energy neutrinos in 110 days. It dominates the neutrino flux in coincidence with the source position over the last 9.5 years for which we have data. It is interesting to note that a subset of blazars, around $1\%-10\%$ of all blazars, bursting once in 10 years at the levels of TXS, can accommodate the diffuse cosmic neutrino flux observed by IceCube. The energy of the neutrino flux potentially generated by the flaring blazars is at the same level as the flux in extragalactic cosmic rays, the Waxman-Bahcall bound.

The coincident observation of Fermi and IceCube, the significance of the 2014 neutrino flare, and the detection of the TeV emission by MAGIC, puts the discovery of the first comic ray accelerator beyond question.

Studying the relation between the diffuse neutrino flux and point source flux of TXS, we derive that the efficiency of the proton beam for producing the pions that are the parents of the observed neutrinos must be close to unity. We will investigate the above facts with detailed modeling of this blazar jet. Early results based on the interaction of the accelerated protons with the dominant 10\,eV blue photons in the galaxy, known as “the blue bump,” are particularly promising.

In summary, we intend to model the rich multiwavelength data provided by this event. The challenge is that the blazar jet must have a sufficiently dense photon target to produce the neutrinos seen by IceCube and, at the same time, be transparent to the TeV photons implied by the MAGIC observation and the large redshift of the source.

\section{Astrophysical modeling to optimize multi-messenger searches}

How effective multi-messenger searches will be in yielding astrophysical insights strongly depends on how well we understand the sources we are searching for, and how much information on these sources we incorporate in searches. Current multi-messenger searches \cite{2016PhRvD..93l2010A,2016ApJ...826L..13A} typically build on minimal source assumptions, primarily operating with a coincidence time window. We will examine the constraints and uncertainties of the emission properties for some of the primary multi-messenger sources \cite{2013RvMP...85.1401A}, and design a strategy that optimizes search sensitivity based on our understanding of the emission. In particular, we will study compact binary mergers and core collapse supernovae with rapidly rotating cores. We will focus on (i) constraints on the progenitor from the GW signal, (ii) relative time of emission, and (iii) relative flux of GWs, neutrinos and electromagnetic counterparts. 

An important goal is to understand the role of a delay between the merger and the jet launching events. The longer the delay between the merger and the jet launching, the larger and more massive is the polluted region \cite{2014ApJ...788L...8M,2017ApJ...850L..24G}. Understanding the consequence of the jet-ejecta interaction on the radiation properties can potentially set constraints on the length of the delay. This can give valuable constraints to GW searches as well as improving our understanding of the launching of relativistic jets from compact binary merger remnants, i.e. newly formed black holes or long-lived supramassive neutron stars.

We expect that the sources associated with GW detections may produce neutrinos in the IceCube energy range.  Moreover these sources are expected to be at are low to intermediate redshift, it is not unreasonable to surmise that meaningful neutrino detections or upper limits will be provided by the IceCube experiment \cite{2003NuPhS.118..388K}. We will estimate the expected relative luminosities of GWs, neutrinos and electromagnetic counterparts for different source models. This will allow multi-messenger searches to use the observed flux in one messenger to calculate a flux prior in another messenger.

\section{ Hidden cosmic accelerators: gravitational wave and neutrino emission scenarios}

We will revise the expected high-energy neutrino flux, along with GW emission models, from hidden sources that are opaque in gamma-rays and therefore are consistent with the IceCube-Fermi constraints. In particular we they are exwill consider transient neutrino sources, choked GRBs \cite{2001PhRvL..87q1102M} and low luminosity GRBs \cite{2006ApJ...651L...5M}. 

We will also consider emission in jets driven by compact binary mergers that interact with sub-relativistic dynamical/wind ejecta from the merger, which can result in attenuated gamma emission prior to the jet burrowing through the slower ejecta, while trans-ejecta neutrinos can escape \cite{2018arXiv180511613K}. 

Merging binary systems containing two compact objects, i.e. a double neutron star, or a stellar-mass black hole and neutron star are hypothesized to be the progenitors of short GRBs.  These mergers are also powerful GW sources within the LIGO sensitive band. 

The ratio of short GRBs to GW events depends on the fraction of mergers that produce gamma-rays, but also on the beaming factor of gamma-ray emission with respect to the more isotropic GW emission. The observational connection between observed GW events and short GRBs therefore can provide constraints on the physical mechanism and nature of the GRB jetted emission.

\section{Conclusions and plan for the future}

The study of common sources of GWs, neutrinos and gamma-rays requires a broad understanding of the emission processes and detection technique. 
It is fundamental that multimessenger observations are performed following a more clear theoretical framework, instead of combined from different studies that have a diverse set of assumptions. 
How effective multimessenger searches will be in yielding astrophysical insights strongly depends on how well we understand the sources we are searching for, and how much information on these sources we incorporate in searches.

The studies described in this paper will build on observations by the Advanced LIGO \cite{2015CQGra..32g4001T}, Advanced Virgo \cite{2015CQGra..32b4001A}, KAGRA \cite{PhysRevD.88.043007}, IceCube \cite{2003NuPhS.118..388K}, ANTARES \cite{2011NIMPA.656...11A}, Fermi \cite{2009ApJ...702..791M,2012ApJS..203....4A} and Swift \cite{2004ESASP.552..777G} detectors, and potentially others, as well as information from future detectors, namely KM3NeT \cite{2006NIMPA.567..457K}, ISS-TAO \cite{2013ExA....36..505C}, and ULTRASAT \cite{2014AJ....147...79S}.

We plan to derive quantitative models for some of the most promising multi-messenger transient sources, namely binary neutron star and neutron star--black hole mergers (presumably short GRBs) and core-collapse supernovae with rapidly rotating cores (long GRBs), delayed neutrino emission from core-collapse supernovae.

We plan to renew the model of hidden or faint gamma-ray sources, in particular choked/low-luminosity/trans-ejecta GRBs. We will consider the possibility of choked short GRBs by examining the prospects of these sources for joint GW+neutrino observations. 

We will also expand on the recent models of neutrino production in jets of binary neutron star mergers \cite{2017ApJ...850L..35A,2018arXiv180511613K,2018MNRAS.476.1191B,2017ApJ...848L...4K}, in particular by studying the effect of the sub-relativistic ejecta and viewing angle on detectability. This scenario became particularly interesting since the multi-messenger discovery of binary neutron star merger GW170817 \cite{PhysRevLett.119.161101,2041-8205-848-2-L12,2017ApJ...850L..35A}. Future observations of similarly nearby mergers whose jets contain a significant fraction of hadrons and that are beamed towards the Earth are expected to be detectable by IceCube for source directions on the northern hemisphere \cite{2018arXiv180511613K}.

We will revise neutrino emission models from these sources and derive their detectability with km-scale neutrino detectors. We will combine this information with the detectability of GWs from the same sources to quantitatively estimate the prospects of multi-messenger observations.

\newpage
\bibliography{References.bib}

\end{document}